\documentclass[aps,pra,twocolumn,showpacs,floatfix,superscriptaddress]{revtex4-1}
\usepackage{graphicx,amsmath,bbm,mathrsfs,amssymb,times}
\begin{document}
\title{Dynamical paths and universality in continuous variables open systems}
\author{Andrea Cazzaniga}
\affiliation{Dipartimento di Fisica, Universit\`a degli Studi di Milano, I-20133,
Milano, Italy}  
\author{Sabrina Maniscalco}
\affiliation{School of Engineering \& Physical Sciences,
Heriot-Watt University Edinburgh, EH144AS, UK}
\affiliation{Turku Centre for Quantum Physics, Department of Physics
and Astronomy, University of Turku, FI-20014 Turun yliopisto,
Finland}
\author{Stefano Olivares}
\affiliation{Dipartimento di Fisica, Universit\`a degli Studi di Milano, I-20133,
Milano, Italy}  
\affiliation{CNISM -- UdR Milano Statale, I- 20133, Milano, Italy}
\author{Matteo G. A. Paris}
\affiliation{Dipartimento di Fisica, Universit\`a degli Studi di Milano, I-20133,
Milano, Italy}  
\affiliation{CNISM -- UdR Milano Statale, I- 20133, Milano, Italy}
\begin{abstract}
We address the dynamics of quantum correlations in continuous variable
open systems and analyze the evolution of bipartite Gaussian states in
independent noisy channels. In particular, upon introducing the notion
of {\em dynamical path} through a suitable parametrization for symmetric
states, we focus attention on phenomena that are common to Markovian and
non-Markovian Gaussian maps under the assumptions of weak coupling and
secular approximation.  We found that the dynamical paths in the
parameter space are universal, that is they do depend only on the
initial state and on the effective temperature of the environment, with
non Markovianity that manifests itself in the velocity of running over a
given path. This phenomenon allows one to map non-Markovian processes
onto Markovian ones and it may reduce the number of parameters needed to
study a dynamical process, e.g. it may be exploited to build constants
of motions valid for both Markovian and non-Markovian maps.
Universality is also observed in the value of Gaussian discord at the
separability threshold, which itself is a function of the sole initial
conditions in the limit of high temperature. We also prove the existence
of excluded regions in the parameter space, i.e. of sets of states which
cannot be linked by any Gaussian dynamical map.
\end{abstract}
\date{\today}
\pacs{03.65.Ta,42.50.Dv,42.50.Xa}
\maketitle
\section{Introduction}
As soon as quantum correlations \cite{reve,revd} have been recognized
as a resource for quantum information processing, it has been realized
that decoherence is the main obstacle to overcome in order to
effectively implement quantum technologies.  Decoherence appears
whenever a system interacts with its environment, so that its dynamics
is no longer unitary, but rather described by a non unitary completely
positive quantum map, irreversibly driving the system towards
relaxation and the loss of quantum coherence \cite{z1,BrePet}.  The
main effect of the interaction with environment is to set up a time
scale $\tau_{\scriptscriptstyle M}\!$ over which the dynamics of the
system is effectively described by a coarse grained Markovian process
towards equilibrium. Conversely, for times shorter than
$\tau_{\scriptscriptstyle M}$, the dynamics is more involved and the
correlations with and within the environment play a major role
\cite{BrePet,Weiss,c1,smi11,nmQJ,smi13}.  In this regime, decoherence may be
less detrimental, and the dynamics may even induce {\em re-coherence}:
this is why a great attention has been devoted to the study of the
corresponding {\em non-Markovian} maps, e.g. in different continuous
variable systems ranging from quantum optics to mechanical oscillators
and harmonic lattices \cite{man07,Paz09,rugg,Cor12,And08}.  Besides,
there are evidences that non-Markovian open quantum systems
\cite{Bre09,m2,m3,NMBreur} can be useful for quantum technologies
\cite{cry11,est11}.  As a consequence, much attention is currently
devoted to the analysis of system-environment coupling in order to
characterize, control, and possibly reduce decoherence in the most
effective way \cite{d1,d2}, e.g. by taking advantage of the back-flow
of information from the environment.
\par
As a matter of fact, non-Markovian models are usually more involved
than the corresponding Markovian ones, and only few cases can be
solved analytically \cite{BrePet,hpz,for01}. However, these cases are
also of great interest, since they display a rich phenomenology which
is relevant for practical implementations. This is especially true for
continuous variable systems \cite{rev12}, where considering a set of
quantum oscillators excited in a Gaussian state, and then linearly
interacting with their thermal environment, provides an excellent
model for a large class of physical systems in order to study
non-Markovianity and the decoherence of quantum correlations.
Motivated by these considerations, we address in details the dynamics
of quantum correlations between two quantum oscillators each
interacting with a local thermal environment. We assume a weak
coupling between the oscillators and the corresponding environment, as
well as the validity of the secular approximation approximation.  These are
the minimal assumptions to have a model that displays remarkable
differences between Markovian and non Markovian dynamics and, at the
same time, allows the use of analytic tools to describe results.  We
also assume the oscillators initially prepared in a symmetric Gaussian
state, and study in details the evolution of their quantum
correlations as described by their dynamical paths, i.e. the time
evolution in a suitably chosen parameter space.
\par
We start by noticing that the set of Gaussian states, i.e. states with
a Gaussian Wigner function \cite{Oli12}, do not constitute a manifold,
nor it is convex, and thus geometrical approaches to their dynamics
are not considered particularly appealing. At variance with this
belief, we address the study of decoherence by representing dynamical
paths in a suitable, overcomplete, parameter space, involving Gaussian
entanglement (negativity) \cite{sim00}, Gaussian discord
\cite{GQD10,Ade10} and the overall purity of the state. The use of
these variables offers a suitable framework to compare non-Markovian
maps and their Markovian counterparts, and to show which properties
do, and {\em notably do not}, distinguish Markovian and non-Markovian
processes. In particular, upon describing the dynamics as a path in
the three-dimensional space individuated by the above variables, we
observe {\em universality}: the dynamical paths do not depend on the
specific features of the environment spectrum and are determined only
by the initial state and the effective temperature of the
environment. The non-Markovianity of the system only changes the
velocity of running over a given path.  This behavior allows one to
map non-Markovian processes onto Markovian ones and it may reduce the
number of parameters needed to study a dynamical process, e.g. it may
be exploited to build constants of motion valid for both Markovian
and non-Markovian maps.  Universality is also observed in the value of
discord at the separability threshold, which moreover is a function of
the sole initial conditions in the limit of high temperature. Finally,
we find that the geometrical constraints provided by the structure of
the parameter space implies the existence of excluded regions,
i.e. sets of states which cannot be linked by any Gaussian dynamical
map.
\par
The paper is structured as follows. In Section II we introduce the
interaction model and briefly review its solution for Gaussian
states. We also introduce symmetric Gaussian states and the quantities
used to quantify their quantum correlations, i.e. Gaussian entanglement
and Gaussian discord. The dynamics of the system is then described in
details in both the Markovian and the non Markovian regimes,
illustrating universality of the dynamical paths. Section III closes the
paper with some concluding discussions and remarks.
\section{Kinematics and dynamics}
Here we consider the dynamical decoherence of two oscillators of
frequency $\omega_0$, each coupled to its own bosonic environment made
of modes at frequencies $\omega_k$. The baths are separated and of
equal structure (see \cite{Paz09,Cor12} for the interaction with a
common bath).  The system-bath interaction Hamiltonian is given by
(we use natural units)
$$H_I = \alpha \sum_k j_k\,(X_1 q_{1k}+X_2 q_{2k})\,,$$
where $\alpha$ is the coupling, the complex $\{j_k\}_{k=1,2..}$
modulate the dispersion over the bath's modes, and $X_s=(a_s
+a^\dag_s)/\sqrt{2}$, $s=1,2$ and $q_{sk}=(b_{sk}+b^{\dag}_{sk}) /\sqrt{2}$
denote the canonical operators of the systems' and baths' modes
respectively, i.e. $[a_s,a^\dag_{s^\prime}]=\delta_{ss^\prime}$,
$s=1,2$ and $[b_k,b_{k^\prime}^{\dag}]=\delta_{kk^\prime}$.  In the
weak coupling limit $\alpha\ll\omega_0$ and performing the secular
approximation 
we can write a non-Markovian time-local master
equation for the dynamical evolution of the density operator $\varrho$
describing the quantum state of the two oscillators in the interaction
picture \cite{hpz}
\begin{align}
\dot\varrho(t)= - \sum_k \Big(& \Delta(t) [X_k,[X_k,\varrho(t)]] \notag
\\ & -i\gamma(t)\,[X_k,\{P_k,\varrho (t)\}] \Big)
\label{eq:HUPAZme}
\end{align}
where $[A,B]$ and $\{A,B\}$ denote commutator and anticommutator between
the operators $A$ and $B$. Upon defining the spectrum of environment as 
$$j(\omega)=\sum_k|j_k|^2\delta(\omega-\omega_k)\,,$$ 
the diffusion (heating) and dissipation (damping) coefficients 
are given by \cite{SabLin}:
\begin{align}
\Delta(t)&=\alpha^2\int_0^t\!\! ds\!\! \int_0^{\infty}\!\!\!\!\!
d\omega\, j(\omega)\coth(\omega\beta/2)\cos(\omega s)\cos(\omega_0s)
\nonumber \\ 
 \gamma(t)&=\alpha^2\int_0^t\!\! ds\!\! \int_0^{\infty}\!\!\!\!\!
d\omega\, j(\omega)
\sin(\omega s)\sin(\omega_0s)\,\label{eq:gamma}\,,
\end{align}
respectively, with $\beta=1/T$.
At high temperatures the damping coefficient $\gamma(t)$ 
is negligible and the diffusion $\Delta(t)$ is dominant, while at 
lower temperatures they have the same order of magnitude.  
\par
It is worth noticing that the assumptions of weak coupling 
and secular approximation are the minimal ones to have a model that displays 
differences between Markovian and non Markovian dynamics.
At the same time the dynamical equations remain simple enough to allow 
the use of analytic tools to describe results.
\par
In fact, the master equation (\ref{eq:HUPAZme}) may be transformed into 
a differential equation for the two-mode symmetrically ordered characteristic 
function associated with the density operator $\varrho$ \cite{man07,RWA}
$$
\chi (\boldsymbol \Lambda) = \hbox{Tr}[\varrho\,
D(\lambda_1)\otimes D(\lambda_2)]
\,,
$$
where $D(\lambda)=\exp\{\lambda a^\dag - \lambda^* a\}$, $\lambda \in
{\mathbb C}$, is the displacement operator and
$\boldsymbol\Lambda = (x_1,y_1,x_2,y_2)$, 
$\lambda_k = (x_k+ i y_k)/\sqrt{2}$.
The solution of this equation
may be written as follows
\begin{align}\label{chit}
\chi_t(\boldsymbol\Lambda)=&
\exp\left\{-\boldsymbol\Lambda^T \left( \widetilde{W}_t\oplus\widetilde{W}_t
\right)\boldsymbol\Lambda\right\}\\
&\times\chi_0\left(e^{-\frac12\Gamma(t)}\left({O}^{-1}_t\oplus
{O}^{-1}_t\right)\boldsymbol\Lambda\right)\,, \notag
\end{align}
where $\chi_t(\boldsymbol\Lambda)$ is the characteristic function 
at time $t$ and $\chi_0(\boldsymbol\Lambda)$ the corresponding quantity
at $t=0$. The quantity $\Gamma(t)$ represents an 
effective time-dependent damping factor, given by
$$\Gamma(t)=\int_0^t\!\!ds\,\gamma(s)\,.$$ The $2\times 2$ matrices
$\widetilde{W}_t$ and $O_t$ are given by
\begin{align}
\widetilde{W}_t=e^{-\Gamma(t)}\, O_t\, W_t\, O^{T}_t\,,
\end{align}
\begin{equation}
O_t=\left(
      \begin{array}{cc}
        \cos\omega_0 t & \sin\omega_0 t \\
        -\sin\omega_0 t & \cos\omega_0 t \\
      \end{array}
    \right).
\end{equation}
Finally, $$W_t=\int_{0}^{t}\!ds\, e^{\Gamma(s)}\widetilde{M}_s$$ 
with
$\widetilde{M}_s=O^{T}_s\,M_s\, O_s$ and
\begin{equation}
\label{eq:Mt} M_s=\left(
       \begin{array}{cc}
         \Delta (s) & 0 \\
         0& 0 \\
       \end{array}
     \right)\,.
\end{equation}
Gaussian states have Gaussian characteristic function and they are
fully characterized by the mean values of the canonical operators
$X_k$ and $P_k=i(a^\dag_k-a_k)/\sqrt{2}$ and by the covariance
matrix (CM) $\sigma$, that writes
\begin{align*}
\sigma_{jk} &= \frac12\,\hbox{Tr}[\varrho (R_j R_k+R_k R_j)],
\end{align*}
with ${\mathbf R} = (X_1,P_1,X_2,P_2)$.  Since the Gaussian character
of an input state is preserved by the master equation
(\ref{eq:HUPAZme}), and we are considering Gaussian states, we need to
address the sole evolution of the first two moments.  In addition, we
can focus attention to the evolution of the CM only, being the quantum
correlations independent of the first moments.
\par
In particular, we assume that the initial state is a two-mode 
Gaussian state $\varrho_0$ with zero
amplitude, i.e. $\hbox{Tr}[\varrho_0\,X_k] = \hbox{Tr}[\varrho_0\,
P_k]=0$, $k=,1,2$, and with covariance matrix $\sigma_0$.
According to Eq. (\ref{chit}) the evolved state at time $t$ is 
still a Gaussian state with zero mean values, and with 
covariance matrix 
given by \cite{rugg,RWA,ale04}
\begin{align}
\sigma_t&=e^{-\Gamma(t)}(O_t\oplus O_t)\sigma_0(O_t\oplus
O_t)^T+\widetilde{W}_t\oplus\widetilde{W}_t\,.
\end{align}
Upon retaining only the terms consistent with the secular 
approximation we arrive at the expression
\begin{equation}
{\sigma}_t=e^{-\Gamma(t)}\sigma_0+ \frac{1}{2}\Delta_{\Gamma}(t)\,
{\mathbb I}_4\label{eq:NMsigma}
\end{equation}
where 
$$
\Delta_{\Gamma}(t)=e^{-\Gamma(t)}\int_0^t\!\!ds\,
e^{\Gamma(s)}\Delta(s)\,,
$$
is a time-dependent effective diffusion factor.
\par
The non-Markovian features are embodied in the time dependence of the
coefficients $\Delta_{\Gamma}(t)$ and $\Gamma(t)$, which describes
diffusion and damping respectively. For times
$t\lesssim\tau_{\scriptscriptstyle M}\!$ both coefficients are strongly
influenced by the whole spectrum of the environment \cite{rugg}.  On the
other hand, for times $t\gg\tau_{\scriptscriptstyle M}\!$ the
coefficients achieve their Markovian limiting values. In particular we
have $$\lim_{t\rightarrow+\infty}
\gamma(t)=\alpha^2|j(\omega_0)|^2\equiv \gamma_{\scriptscriptstyle M},$$
such that 
$$\Gamma(t)=\gamma_{\scriptscriptstyle M}t \qquad 
\Delta_{\Gamma}(t) = (1-e^{- \gamma_{\scriptscriptstyle M}t})(2
n_{\scriptscriptstyle T}+1),$$ and the solution (\ref{eq:NMsigma})
rewrites as
$$\sigma(t)=
e^{- \gamma_{\scriptscriptstyle M}t}
\sigma_0+ (1-e^{- \gamma_{\scriptscriptstyle M}t})
\sigma_{\scriptscriptstyle T},$$ where $\sigma_{\scriptscriptstyle T} =
(n_{\scriptscriptstyle T}+\frac12) {\mathbb I}_4$ is the CM of the
stationary state, i.e. a thermal equilibrium state at temperature
$1/\beta$ and, in turn, a population of $n_{\scriptscriptstyle
T}=(e^{\beta\omega}-1)^{-1}$ thermal photons.
\subsection{Symmetric Gaussian states}
A bipartite Gaussian state is symmetric if its CM can be recast (via
local operations) in a form depending on two real parameters $a$ and
$c$, that is
\begin{align}\label{sym}
\sigma=a\, {\mathbb I}_4 + c\, \sigma_1 \otimes \sigma_3\,, 
\end{align}
the $\sigma_j$'s being Pauli matrices. Note that uncertainty relations
impose a constraint which reads \cite{sim87}
$|a-c|\geqslant\frac{1}{2}$.  The evolution under the master equation
(\ref{eq:HUPAZme}) preserves the symmetry [see
Eq.~(\ref{eq:NMsigma})], therefore the evolved CM at time $t$ may be
still written as $\sigma(t)=a(t)\, {\mathbb I}_4 + c(t)\, \sigma_1
\otimes \sigma_3$, where
\begin{align}
a(t) &=a_0\, e^{-\Gamma(t)}  + \Delta_\Gamma (t)  \\ 
c(t) &=c_0\, e^{-\Gamma(t)}\,,
\end{align}
with $a_0=a(0)$ and $c_0=c(0)$.
\par
A symmetric CM of the form (\ref{sym}) corresponds to prepare the two
oscillators in a {\em squeezed thermal state} (STS), i.e a state with
density operator of the form $$\varrho(r,\nu_{\scriptscriptstyle
  T})=S(r)\,\nu\otimes\nu \,S^{\dag}(r)\,,$$ where $\nu$ denotes a
single mode thermal state with $\nu_{\scriptscriptstyle T}\!$ photons
and $S(r)=e^{r(a_1 a_2 -a_1^\dag a_2^\dag)}$ is the two-mode squeezing
operator. For $\nu_{\scriptscriptstyle T}=0$ the state $\varrho(r,0)$
reduces to the so-called two-mode squeezed vacuum state or {\em
  twin-beam} state, i.e. the maximally entangled state of two
oscillators at fixed energy.
\par
The parameters of the CM are connected to the physical parameters as
follows
\begin{align*}
a=(\nu_{\scriptscriptstyle T}+\frac{1}{2})\cosh(2r) \quad 
c=(\nu_{\scriptscriptstyle T}+\frac{1}{2})\sinh(2r)\,.
\end{align*}
Furthermore, by introducing the (equal) population (mean photon
number) of the two subsystems $$\bar n = \sinh^2 r\, (2
\nu_{\scriptscriptstyle T}+1) + \nu_{\scriptscriptstyle T}\,,$$ the
diagonal elements may be written as $a=\frac{1}{2}+\bar n$, while the
$c$ coefficients describe the correlations among them. It is worth
noting that {\it any} two-mode entangled Gaussian state can be converted
into a symmetric one by local operations and classical communication
\cite{sym1,sym2}. Therefore, our results about the dynamics of quantum
correlations actually holds for more general initial states than the
symmetric ones, including any initially entangled state.
\par
Indeed, the representation in terms of the coefficients $a$ and $c$
does not fully illustrate the correlation properties of a state.  In
particular, it does not allow to analyze the relations between
different kinds of quantum correlations, as entanglement or discord,
in a dynamical context, and to compare their robustness against
dissipation and noise.  To this aim we introduce a different
(overcomplete) parametrization involving the overall purity of the
state
\begin{align}
\mu=\hbox{Tr}[\varrho (r,\nu_{\scriptscriptstyle T})^2]
=\frac{1}{4\sqrt{\det \sigma}}=\frac{1}{(2\nu_{\scriptscriptstyle
T}+1)^2}\,,
\end{align}
its Gaussian entanglement expressed in terms of the minimum symplectic
eigenvalue
$$
\lambda=a-c=(\nu_{\scriptscriptstyle T}+\frac{1}{2})e^{-2r}\,
$$ (the state is
separable iff $\lambda \geq \frac12$) and the Gaussian quantum
discord, which for symmetric Gaussian states may be written as
\cite{GQD10}
\begin{align}\label{eq:discord}
D(a,c) &= h(a)-2h\left(\sqrt{a^2-c^2}\right)+h\left(a-\frac{2c^2}{1+2a}\right) 
\notag \\ &\equiv D(\mu,\lambda)
\end{align}
where
$$h(x)=\left(x+\frac{1}{2}\right)\log\left(x+\frac{1}{2}\right)-
\left(x-\frac{1}{2}\right)\log\left(x-\frac{1}{2}\right)\,.$$
The parameter space individuated by $\mu$, $\lambda$, and $D$ is overcomplete and
the third parameter is a function of the other two \cite{a1} at any
time.  In the following, we will describe the dynamics of the system by paths 
in the three-dimensional space $(\mu,\lambda,D)$ according to the
following definition: 
\\ $ $ \par
{\bf Definition} (Dynamical path): 
{\em a dynamical path for a symmetric Gaussian state is a line 
in the three-dimensional space $(\mu,\lambda,D)$ individuated by the
overall purity of the state $\mu$, its least symplectic eigenvalue
$\lambda$, and its Gaussian discord $D$.}
\\ $ $ \par
Dynamical paths lay
on the surface individuated by the constraint (\ref{eq:discord}) and in the
region satisfying the uncertainty relations. In terms of the parameters 
$(\mu,\lambda,D)$ these constraints correspond to
\begin{align}
D=D(\mu,\lambda) \qquad \mu<\frac{1}{4\lambda^2}\,.
\end{align}
A dynamical path describes the evolution of a symmetric Gaussian state
in a noisy Gaussian
channel with no explicit dependence on time. This allows one to compare 
non-Markovian maps and their Markovian counterparts, and to show which 
properties do, and do not, distinguish 
Markovian and non- Markovian processes. At the same time, it
allows us to reveal the relationships among the different kinds of
quantum correlations in a dynamical context. In other words, each
dynamical path actually describes an equivalence 
class of dynamical time-dependent trajectories (including both Markovian and 
non-Markovian ones), characterized by a specific dependence of the
Gaussian discord on the other two parameters.
\subsection{Markovian dynamics}
The Markovian master equation depends on the (effective) environment
temperature and on the damping $\gamma_{\scriptscriptstyle M}\!$,
nonetheless the Markovian dynamical paths depend exclusively on the
(effective) temperature of the environment.  The damping only affects
the speed of running over a dynamical path, but not its shape, and the
rate $c(t)/c_0=e^{-\gamma_M t}$ determines in a unique way the rate
$a(t)/a_0$.  In the left panel of Fig.~\ref{f:path} we show few
Markovian paths for different values of the temperature, assuming the
two oscillators initially prepared in a two-mode squeezed vacuum (TWB)
$\varrho(r_0,0)$, i.e a pure maximally entangled state of the two
oscillators.  As it is apparent from the plot, two limiting paths
emerge at low and high temperatures.  The transition from one regime
to the other occurs continuously by raising the temperature, and we
see that the high temperatures limit is already achieved for
temperatures corresponding to $$n_{\scriptscriptstyle
  T}/\sinh^2(r_0)\gtrsim 3\,.$$ Two other phenomena are revealed by
this representation: (i) the value of the discord at the separability
threshold ($\lambda=\frac12$) depends only on the initial squeezing
$r_0$ and approaches a universal curve in the high temperature limit;
(ii) for a given initial state $\varrho(r_0,0)$ there are STS that
cannot be reached during any Markovian decoherence process, despite
the fact that they have reduced entanglement and purity compared to
the initial state.
\subsection{Non Markovian dynamics}
As mentioned in the introduction, non-Markovian dynamics may display
remarkable differences from their Markovian counterpart during the
initial transient when $t\lesssim\tau_{\scriptscriptstyle M}$.
Entanglement oscillations may occur, and the separability threshold may
be delayed or accelerated depending on the spectrum of the environment.
A question thus arises on whether these differences also affect
significantly the dynamical path in the space of parameters. As we will
see, the answer is negative, and universality occurs. The results about
the dynamics that we are going to discuss are independent of the
particular choice of environment spectrum, and this is a crucial point
of our analysis. However, in order to show
some numerical solutions, we employ few examples corresponding to white
noise and to both Ohmic and super-Ohmic spectral  densities with cut-off
$\omega_c$.  More specifically, we are going to consider Ohmic spectrum 
$$j(\omega)\propto
\frac{\omega\omega_c^2}{\omega^2+\omega_c^2}\,,\quad \mbox{(Ohmic)}$$ 
which leads to non-Markovian features when out of resonance, i.e. 
when $\omega_0\gg\omega_c$, SuperOhmic spectrum $$j(\omega)\propto
\frac{\omega^2\omega_c}{\omega^2+\omega_c^2}\,\quad \mbox{(super-Ohmic)}$$ 
and white noise spectrum $j(\omega) \propto \omega_c$.
\par
\begin{figure}[h!]
\includegraphics[width=0.42\columnwidth]{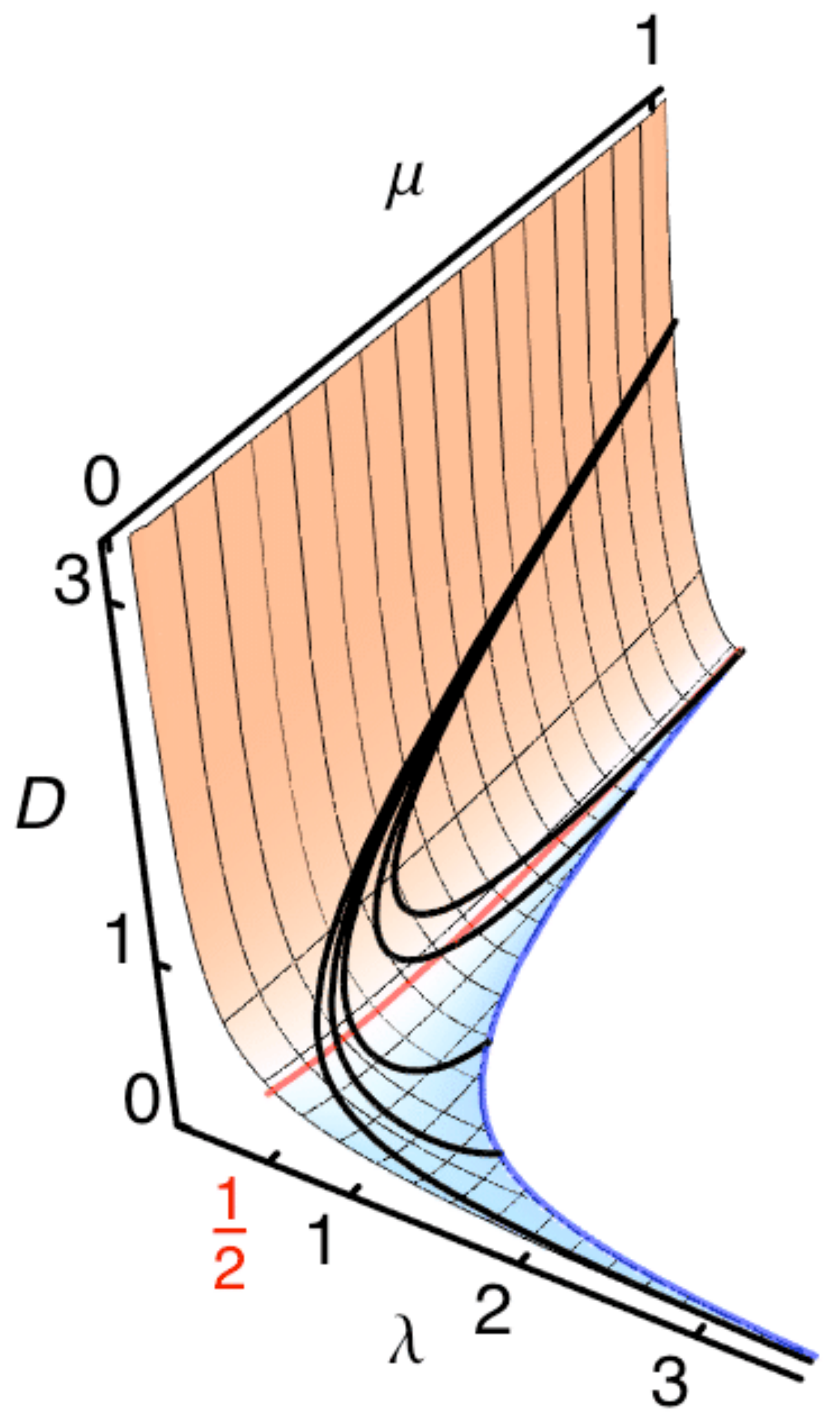}
\includegraphics[width=0.42\columnwidth]{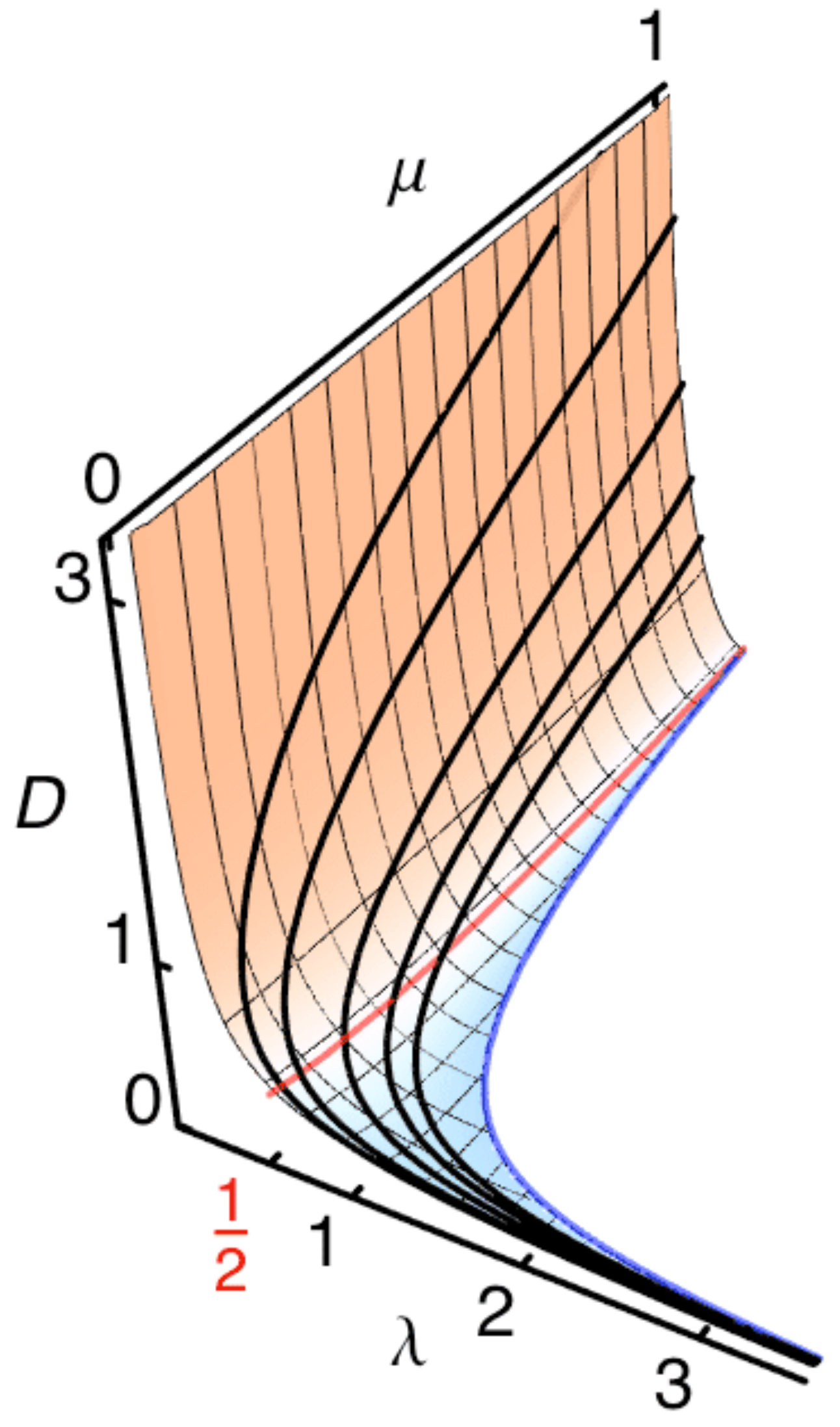}
\caption{(Color online) Left: Markovian thermalization paths at
different temperatures. The initial state of the two oscillators is a
two-mode squeezed vacuum with $r_0=1.2$, the different paths (black
lines) correspond to different number of thermal photons in the
environment. From top to bottom we have paths for $n_{\scriptscriptstyle
T}=0,0.1,0.5,1.0,10$. The solid blue line corresponds to thermal product 
states $\nu\otimes\nu$ with
zero discord. The solid red line denotes the separability threshold.  The
high-temperature limit is already achieved for $n_{\scriptscriptstyle T}
\gtrsim 3$.  Right: dynamical paths in the high-temperatures limit for
the two oscillators initially prepared in a two-mode squeezed vacuum
state with different values of initial entanglement. The curves
correspond to (from bottom to top) $r_0=0.5 ,0.7 ,1.0, 1.5, 2.0$.
\label{f:path}}
\end{figure}

Let us start by analyzing the high temperature regime, where 
over a timescale $\tau\sim\tau_{\scriptscriptstyle M}\!$ we can neglect the 
damping $\Gamma(t)$ (it becomes relevant over times 
$\tau\sim\gamma_{\scriptscriptstyle M}^{-1}\gg\tau_{\scriptscriptstyle M}$, 
which is definitely
in the Markovian regime). Short time non-Markovian dynamics is thus due 
to the behavior of the heating function $\Delta_{\Gamma}(t)$ and, in
turn, is very sensitive to the details of the environment spectrum $j(\omega)$. 
In this limit non-Markovian effects can be seen during the 
whole decoherence process, with entanglement oscillation 
across the separability threshold \cite{man07}. The 
dynamics is driven by the approximate dynamical equation
\begin{equation}
\sigma_t\simeq\sigma_0+\int_0^t
\!\!ds\,\Delta(s)\,\frac{\mathbb{I}_4}{2}
\label{eq:HTmastereq}
\end{equation}
corresponding to
\begin{align}
a(t)&=a_0+\frac{1}{2}\int_0^td\tau\,\Delta(\tau) \\ 
c(t)& =c_0 \,.
\end{align}
The minimum symplectic eigenvalue is thus given by
\begin{align}
\lambda(t)=\lambda_0+\frac{1}{2}\int_0^td\tau\,\Delta(\tau) \,.
\end{align}
The condition $c(t)=c_0$ imposes a constraint to the dynamical
paths, which is the same independently on whether the dynamic of $a(t)$ is 
Markovian or displays oscillations, as long as $a(t)\geq a_0\:\:\forall t$ 
and $a(t)\rightarrow a_{\scriptscriptstyle T}$.  In other words, the paths
are the same of the Markovian case, and the possible oscillations of $a(t)$ 
only influences the speed of running over the dynamical path.   
In the right panel of Fig.~\ref{f:path} we show the dynamical 
paths for different values of the initial squeezing $r_0$.
\subsection{Discord at the separability threshold}
\begin{figure}[ht]
 \includegraphics[width=0.95\columnwidth]{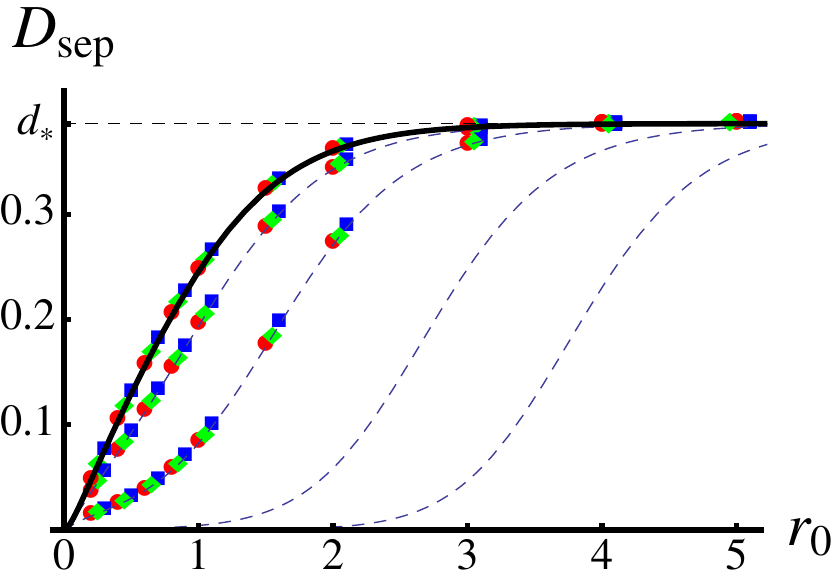}
 \caption{Discord at separability threshold as a function of the initial
 squeezing. The solid black line denotes the universal function 
 $D_{sep}(r_0)$ of Eq. (\ref{dsep}) obtained in the high temperature limit, 
 whereas the colored symbols are the solutions of the 
 full non-Markovian dynamics in Eq.~(\ref{eq:NMsigma}) for
 $n_{\scriptscriptstyle T}=10$ 
 and for three
 different environment spectra corresponding to Ohmic, super-Ohmic, 
 and white noise spectral density respectively (red circles, green
 diamonds, and blue squares). The horizontal dashed line
 is the high-temperature high-squeezing limiting value $d_*
 \simeq0.3863$. The dashed gray lines denote the low temperature 
 Markovian curves
 $D(\frac12+c(t_{sep}), c(t_{sep}))$ for  
 $n_{\scriptscriptstyle T}=0.5, 0.1, 10^{-2}, 10^{-3}$ 
 respectively. We also report the solutions of the 
 full non-Markovian dynamics for  
 $n_{\scriptscriptstyle T}=0.5, 0.1$ 
 and the same three spectra, whereas for 
 $n_{\scriptscriptstyle T}=10^{-2}, 10^{-3}$ the separability 
 threshold is definitely in the Markovian regime.
 \label{f:HTDst}}
\end{figure}

The condition $c(t)=c_0$ also implies that the Gaussian discord may
be written as $$D(a,c)\simeq D(\lambda+c_0,c_0)\qquad T\gg 1\,,$$ i.e. it depends 
on the temperature and on the initial squeezing \cite{isa12} only
through the minimum symplectic eigenvalues. At the separability 
threshold, i.e. for $t=t_{sep}$ such that $\lambda(t_{sep})$, 
we have 
\begin{align}
D_{sep} &\equiv D_{sep} (r_0) \notag \\ 
&= D\left(\frac12 [1+\sinh(2 r_0)],
\sinh (2 r_0)\right)\,,
\label{dsep}
\end{align}
i.e. the discord at separability is a universal function of the initial
squeezing.  In Fig.\ref{f:HTDst} we show the Gaussian discord at 
separability as a function the initial squeezing. The solid black line 
correspond to the above high temperature approximation $D_{sep}(r_0)$, 
whereas the colored symbols correspond to the full
non-Markovian solutions for $n_{\scriptscriptstyle T}=10$, 
obtained taking into account the damping 
and different environment spectra. As it is apparent from the plot, 
there is an excellent agreement among the two 
solutions, independently on the environment spectrum. We also notice
that $D_{sep}$ saturates to a limiting value 
$$d_*= \lim_{r_0\rightarrow \infty} D_{sep}(r_0)
= -1+2 \log 2 \simeq 0.3863$$ as far as the initial squeezing
increases. The initial squeezing needed for achieve the saturation
regime increases with temperature. As it may be seen from the plot, 
for high temperatures, i.e. for $n_{\scriptscriptstyle T}\gtrsim
1$,  it is about $r_0\simeq 2$.
\par
For lower temperature the approximation $c(t)\simeq c_0$ is 
no longer valid and the Gaussian discord at separability is given by
(Markovian expression)
$$D_{sep} \equiv D_{sep}(r_0, n_{\scriptscriptstyle T})
= D(\frac12+c(t_{sep}), c(t_{sep}))\,.$$ 
In Fig. \ref{f:HTDst} we show $D_{sep}(r_0, n_{\scriptscriptstyle T}$
as a function of $r_0$ for different values of $n_{\scriptscriptstyle T}$
(dashed gray lines). We also report the values obtained from 
the full non-Markovian solutions for different environment spectra
and {\em not so low} temperature, i.e.
 $n_{\scriptscriptstyle T}=0.5, 0.1$.
As it is apparent from the plot the two solutions are in excellent
agreement and this may be understood as follows.
At low temperatures the damping $\gamma(t)$ and the heating function
$\Delta(t)$ become of the same order of magnitude and thus the 
separability threshold $t_{sep}$ does depend on the environment
spectrum. On the other hand, separability
is always achieved in the Markovian regime, and 
thus $D_{sep}$ is a universal quantity. 
The plot confirms that this argument holds also if the temperature is
not so low, i.e. for $n_{\scriptscriptstyle T}=0.5, 0.1$.
For times
$t\lesssim\tau_{\scriptscriptstyle M}$, there is a competition between
$\gamma(t)$ and $\Delta(t)$ and in principle, one would not 
expect a universal behavior. However, low temperature and
weak coupling make the effect of damping and heating very weak, with 
appreciable perturbation of the initial state only after a long time. 
In other words, any dynamical effect of the interaction is taking place 
in the Markovian regime, thus re-gaining universality and independence 
on the  environment spectrum. This also means that the dynamical paths 
in the left panel of Fig. \ref{f:path} legitimately describe 
non-Markovian dynamical trajectories at low temperatures.
\subsection{Universality of constants of motion}
Any path-dependent property may be checked analytically using 
the set of Markovian equations and then  extended to the non-Markovian 
regime,  where an analytic 
approach would be unfeasible. In particular, let us introduce 
the rescaled time $\tau=\Gamma t$, and recall that in  
the Markovian regime we have
\begin{align*}
\partial_\tau\lambda & = e^{-\tau} (\lambda_{\scriptscriptstyle
T}-\lambda_0)\\ 
\partial_\tau(\lambda \mu)^{-1}&=
e^{-\tau}\left[(\lambda_0\mu_0)^{-1}+4 \lambda_{\scriptscriptstyle
T}\right]\,,
\end{align*}
where pedices $0/T$ refer to initial/stationary state. Then, any
constant of motion, e.g.
$C=\lambda+y/(4\lambda\mu)$, with $ 
y=(\lambda_{\scriptscriptstyle
T}-\lambda_0)/\lambda_{\scriptscriptstyle
T}+(4\mu_0\lambda_0)^{-1}$
built using the Markovian dynamical equation is 
a constant of motion also in the non-Markovian regime, independently on
the environment spectrum, and with potential application
for the development of general channel engineering strategies.
The temperature dependence disappears in the
high-temperatures limit. 
\section{Discussion and conclusions} 
We have addressed the dynamics of quantum correlations in continuous
variable open systems and analyzed the evolution of bipartite Gaussian
states in independent noisy channels. We have assumed weak coupling
between the system and the environment, as well as the secular
approximation. These are the
minimal assumptions to have a model that displays remarkable
differences between Markovian and non-Markovian dynamics and, at the
same time, allows the use of analytic tools to describe results.
\par
In describing the noisy evolution of two-mode symmetric Gaussian states
we introduced the concept of dynamical paths,  i.e. lines in the
three-dimensional space individuated involving Gaussian entanglement,
Gaussian discord and the overall purity of the state. Dynamical paths
describe the evolution of symmetric Gaussian states with no explicit
dependence on time. This has been proven suitable to address the
decoherence effects of both Markovian and non-Markovian Gaussian maps,
and to reveal which properties do, and  do not, distinguish Markovian
and non-Markovian processes.  At the same time, dynamical paths allow us
to reveal the relationships among the different kinds of quantum
correlations in a dynamical context.  Each dynamical path actually
describes an equivalence class of dynamical time-dependent trajectories
(including both Markovian and non-Markovian ones), characterized by a
specific dependence of the Gaussian discord on the other two parameters.
\par
Upon describing the dynamics as a
path in the three-dimensional space individuated by the above variables,
we have observed universality: The dynamical paths do not depend on the
specific features of the environment spectrum and are determined only by
the initial state and the effective temperature of the environment.
Non Markovianity manifests itself in the 
velocity of running over a given path. This phenomenon allows one to map 
non-Markovian processes onto Markovian ones and it may reduce the number 
of parameters needed to study a dynamical process, e.g. it may be exploited 
to build constants of motions valid for both Markovian and non-Markovian maps.
\par
Universality is also observed for the value of discord at the
separability threshold, which moreover depends on the sole initial
squeezing in the high temperature limit.  We also found that the
geometrical constraints provided by the structure of the parameter space
implies the existence of excluded regions, i.e. sets of Gaussian states
which cannot be linked by any Gaussian dynamical map, despite the fact
that they have reduced entanglement and purity compared to the initial
one.  
\par
Our results have been obtained for Gaussian states and are not 
directly transferable to the non Gaussian sector of the Hilbert space.
Indeed, there are no necessary and sufficient criteria to individuate and 
quantify non Gaussian entanglement, and there are no analytic formulas 
to evaluate non Gaussian quantum discord. The interplay between Gaussian 
and non Gaussian quantum correlations has been discussed in the recent 
years \cite{nGL,ng1,ng2,ng3}, but a complete understanding has not yet 
been achieved. 
\par
Finally, we emphasize once again that the universality of dynamical paths 
does not depend on the environment spectrum, i.e. it is a consequence
of the sole assumptions of weak coupling and the linear interaction 
between system and environment. It may therefore be 
conjectured that universality represents a more general feature, 
characterizing any open quantum system admitting a Markovian limit.
\acknowledgments
This work has been supported by MIUR (FIRB LiCHIS-RBFR10YQ3H), EPSRC
(EP/J016349/1), the Finnish Cultural Foundation ({\em Science Workshop on
Entanglement}), the Emil Aaltonen foundation ({\em Non-Markovian Quantum
Information}) and SUPA. SO, and MGAP thanks Ruggero Vasile for useful
discussions. 

\end{document}